\documentclass[conference]{IEEEtran}
\IEEEoverridecommandlockouts

\usepackage{booktabs}
\usepackage{multirow}
\usepackage{subcaption}
\usepackage[dvipsnames]{xcolor}
\usepackage{url}
\usepackage{graphicx}
\usepackage{hyperref}
\usepackage{stix}
\usepackage{tcolorbox}
\usepackage{listings}
\usepackage{tcolorbox}
\tcbuselibrary{breakable}
\usepackage{tabularx}
\usepackage{array}
\usepackage{makecell}
\usepackage{algorithm}
\usepackage{algorithmic}

\AtBeginDocument{%
  }

\begin{document}

\title{PyGress: Tool for Analyzing the Progression of Code Proficiency in Python OSS Projects}

\author{Rujiphart Charatvaraphan\textsuperscript{*},
  Bunradar Chatchaiyadech\textsuperscript{*},
  Thitirat Sukijprasert\textsuperscript{*}, 
  Chaiyong Ragkhitwetsagul\textsuperscript{*},\\
  Morakot Choetkiertikul\textsuperscript{*},
  Raula Gaikovina Kula\textsuperscript{\textdagger},
  Thanwadee Sunetnanta\textsuperscript{*}, 
  Kenichi Matsumoto\textsuperscript{$\ddagger$}\\
  \textsuperscript{*}\textit{Faculty of Information and
  Communication Technology, Mahidol University, Thailand}\\
  \textsuperscript{\textdagger}\textit{Graduate School of Information Science and Technology, The University of Osaka, Japan}\\
  \textsuperscript{$\ddagger$}\textit{Graduate School of Information Science, Nara Institute of Science and Technology, Japan}}

\maketitle

\begin{abstract}
Assessing developer proficiency in open-source software (OSS) projects is essential for understanding project dynamics, especially for expertise. This paper presents ``PyGress'', a web-based tool designed to automatically evaluate and visualize Python code proficiency using pycefr, a Python code proficiency analyzer. By submitting a GitHub repository link, the system extracts commit histories, analyzes source code proficiency across CEFR-aligned levels (A1–C2), and generates visual summaries of individual and project-wide proficiency. The PyGress tool visualizes per-contributor proficiency distribution and tracks project code proficiency progression over time. PyGress offers an interactive way to explore contributor coding levels in Python OSS repositories.
The video demonstration of the PyGress tool can be found at \url{https://youtu.be/hxoeK-ggcWk}, and the source code of the tool is publicly available at \url{https://github.com/MUICT-SERU/PyGress}.
\end{abstract}


\maketitle

\section{Introduction}
\label{sec:introduction}
Python, known for its ease of use and adaptability to various domains, stands as a versatile and most popular language~\cite{github_octoverse_2024} with a plethora of popular open-source software (OSS) projects. 
Python's simplicity in coding empowers developers to swiftly prototype and construct applications. What sets Python apart is its rich ecosystem, encompassing a multitude of OSS libraries that cater to diverse needs. Libraries such as \textsf{NumPy}, \textsf{Pandas}, and \textsf{Scikit-learn} are pivotal in the fields of data science and machine learning. \textsf{Matplotlib}, \textsf{Plotly}, and \textsf{Seaborn} are used for static plots and data representations. 

OSS projects thrive on contributions from developers with diverse backgrounds and varying levels of expertise. Understanding contributors’ proficiency can help uncover patterns of project success, contributor retention, and growth. %
A distinctive characteristic of OSS is its emphasis on transparency.
In contrast to commercial projects, open-source initiatives embrace a collaborative development model, distributed control, and a diverse contributor base. 
The survival of open-source projects hinges on factors like robust community support, dedicated maintainer commitment, adequate funding, and ongoing relevance. 
The bus factor~\cite{busfactor}, representing a project's vulnerability to knowledge loss caused by losing key project contributors, is also a concern in OSS, urging practices such as documentation, diverse contributions, and shared responsibilities. Thus, OSS maintainers must focus on fostering a welcoming community for new and existing contributors, and ensuring the long-term sustainability of the project, often within a volunteer-based environment with flexible deadlines~\cite{Dias2021}.

To effectively ensure the expertise required to maintain an OSS project, our key idea lies in assessing developers' proficiency, i.e., an individual's programming skills. 
This encompasses the capability to not only write code but also understand and effectively debug it. %
By knowing the contributor's proficiency, maintainers can guide and mentor new contributors, which includes delegating suitable coding tasks for them~\cite{Dias2021}.

\begin{figure}[tb]
    \centering
    \includegraphics[width=0.85\columnwidth]{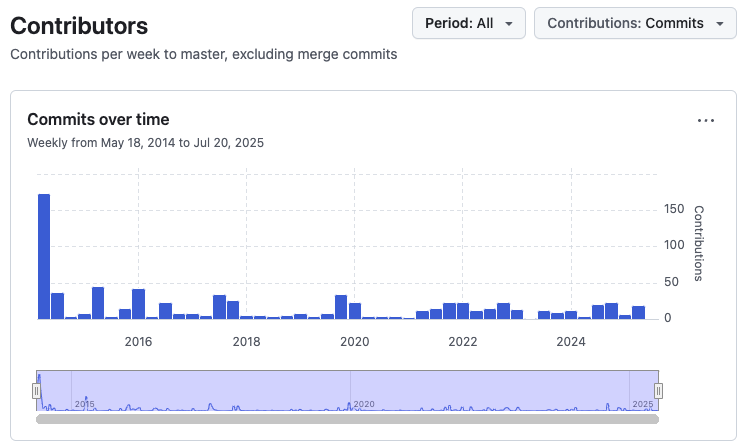}
    \caption{ Current state-of-the-art only depicts contributions over time. This example is from the \textsf{Django-silk} Project}
    \label{fig:github_contributors}
\end{figure}

Currently, there is a noticeable gap in the Python OSS ecosystem—\textit{there is no automated way available for measuring and visualizing proficiency levels over time, which is crucial for tracking contributions in OSS projects.} 
OSS project maintainers lack a reliable method to differentiate contributors based on their code proficiencies, hindering the ability to understand the skill distribution within the teams. %
For example, Figure~\ref{fig:github_contributors} shows the contributors and their code contributions to the \textsf{django-silk}\footnote{https://github.com/jazzband/django-silk} project, a live profiling and inspection tool for the Django framework, on GitHub. With this information, the maintainers of the project only know the amount of code committed by each contributor over time. However, this contribution summary merely shows the frequency of the commits, and the maintainers do not gain any insights into the proficiency levels of such committed code.
 
To address this, we introduce \textit{\textbf{PyGress}, a web-based prototype tool for automated analysis and visualization of developer proficiency in Python OSS projects.} PyGress assesses Python source code against the Common European Framework of Reference (CEFR)~\cite{cefr} proficiency levels, ranging from A1 (basic) to C2 (proficient).
Our automated tool accepts a GitHub repository URL and evaluates Python code from each commit, and generates visualizations that support both project-wide and contributor-specific insights.

\section{Background}
\subsection{The Common European Framework of Reference for Languages (CEFR)}
\label{sec:cefr}
CEFR~\cite{cefr} is a globally recognized standard for evaluating language proficiency, which categorizes language skills using a six-level scale ranging from A1 (breakthrough or beginner), A2 (waystage or elementary), B1 (threshold or intermediate), B2 (vantage or upper intermediate), C1 (effective operational proficiency or advanced), and C2 (mastery or proficiency). This system simplifies the assessment of language competence for educators, students, and others involved in language education and testing. In addition, the CEFR also facilitates the comparison of qualifications with other national examinations, benefiting employers and academic institutions.

\subsection{pycefr: Python Proficiency Level through Code Analysis}
pycefr~\cite{Robles2022} is an automated tool to assess proficiency in Python projects. It analyzes a given Python project and classifies Python constructs into one of the six levels mirroring the CEFR framework. For example, an \texttt{if} and nested list statements are classified as A1 and A2 (basic), respectively. \texttt{break} and list comprehension are classified as B1 and B2 (intermediate). Lastly, the generator function and metaclass are classified as C1 and C2 (proficient), respectively. We adopt pycefr as a key component in PyGress to analyze the proficiency of OSS developers and projects over multiple commits.

\section{PyGress: Evolution of Code Proficiency}

PyGress enables the analysis of Python code proficiency evolution within open-source projects. This is achieved by examining the complete commit history of a given GitHub repository. For each commit, the tool extracts the Python source files and analyzes them using pycefr to determine proficiency levels aligned with the six CEFR levels.

\begin{algorithm}[tb]
\caption{Extracting and Aggregating Contributor's Proficiency Scores}\label{alg:extract}
\footnotesize
\textbf{Input:} project \\
\textbf{Output:} proficiencies \\
\textbf{Procedure:}~ExtractContributorProficiencies
\begin{algorithmic}[1]
    \STATE $\mathrm{proficiencies} \gets \{\}$
    \STATE $\mathrm{committers} \gets \{\}$
    \STATE $\mathrm{commits} \gets \mathrm{extract\_commits(\mathrm{project})} $
    \FOR {$ c_i \in \mathrm{commits}$}
        \STATE $\mathrm{committer} \gets \mathrm{get\_committer(c_i)} $
        \STATE $\mathrm{add\_commits}(\mathrm{committer},c_i)$
        \STATE $\mathrm{committers} \cup \mathrm{committer}$
    \ENDFOR
    \FOR {$ \mathrm{committer} \in \mathrm{committers}$}
        \FOR {$ c_i \in \mathrm{get\_all\_commits(\mathrm{committer})}$}
        \FOR {$ f_i \in \mathrm{changed\_files(c_i)}$}
            
            \STATE $\mathrm{before} \gets \mathrm{get\_version(f_i,c_{i-1})} $
            \STATE $\mathrm{after} \gets f_i$
            \STATE $p_{f_i,\mathrm{before}} \gets \mathrm{pycefr(before)}$
            \STATE $p_{f_i,\mathrm{after}} \gets \mathrm{pycefr(after)}$
            \STATE $\mathrm{proficiency}_{\mathrm{committer}, c_i} \gets p_{f_i,\mathrm{after}} - p_{f_i,\mathrm{before}}$
            \STATE $\mathrm{proficiencies} \cup \mathrm{proficiency}_{\mathrm{committer}, c_i} $
        \ENDFOR
        \ENDFOR
    \ENDFOR
\end{algorithmic}
\end{algorithm}

The steps PyGress uses to extract and aggregate the contributors' proficiency scores of any Python projects are described in Algorithm~\ref{alg:extract}. First, PyGress collects all the commits in each project and groups them by the committers (lines 4--8). Then, it iterates through each committer, retrieves the commits, and extracts all the changed Python files in that commit (lines 9--11). Next, PyGress extracts the file \textit{before} and \textit{after} the commit (lines 12--13). The complete Python files before and after the commit are required because pycefr works at the file level. Then, pycefr is executed by giving the before and after versions of the same Python file, and the proficiency of the code introduced in the commit is collected by computing the differences between the proficiency scores of the after and before versions (lines 14--16). Lastly, the derived proficiency scores are aggregated into the set of project proficiencies over all the commits (line 17).  

\begin{figure}[tb]             
    \includegraphics[width=0.9\columnwidth]{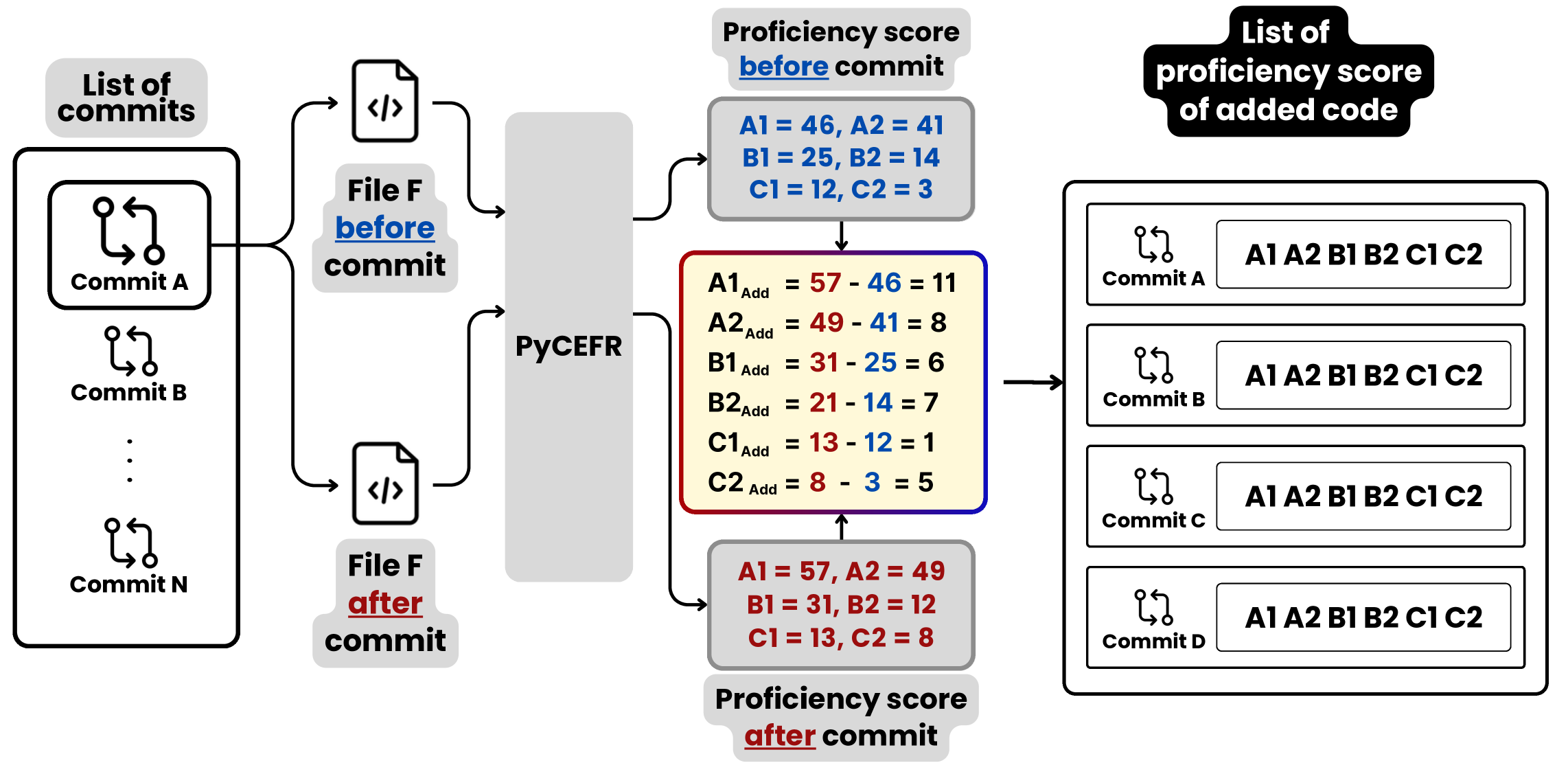}
    \caption{Approach for analyzing code proficiency changes}
    \label{fig:changes_proficiency}
\end{figure}

In Figure~\ref{fig:changes_proficiency}, we illustrate further the process of analyzing the proficiency of the code introduced in the commit, i.e., \textit{added proficiency scores}. After getting the Python code before and after the commit, PyGress subtracts the frequency of code proficiency before the commit from the score after the commit, in each level. For example, if the proficiency score (\{\textit{A1, A2, B1, B2, C1, C2}\}) of a Python file before a commit is \{\textit{46, 41, 25, 14, 12, 3}\} and after the commit is \{\textit{57, 49, 31, 12, 13, 8}\}. Then, this commit introduces 38 new code constructs classified into the six-CEFR levels as \{\textit{11, 8, 6, 7, 1, 5}\} accordingly. We discarded negative scores caused by code deletion, i.e., replaced them with 0, as we focus only on the added code.
This score is added to the pool of proficiency scores for that contributor.

The concept of PyGress can be applied to other programming languages by switching the code proficiency level from pycefr to others (e.g., jscefr~\cite{Ragkhitwetsagul2024} for JavaScript).

\section{Tool Implementation}
\subsection{System Architecture Design}
\begin{figure}[tb]
    \centering
    \includegraphics[width=0.95\columnwidth]{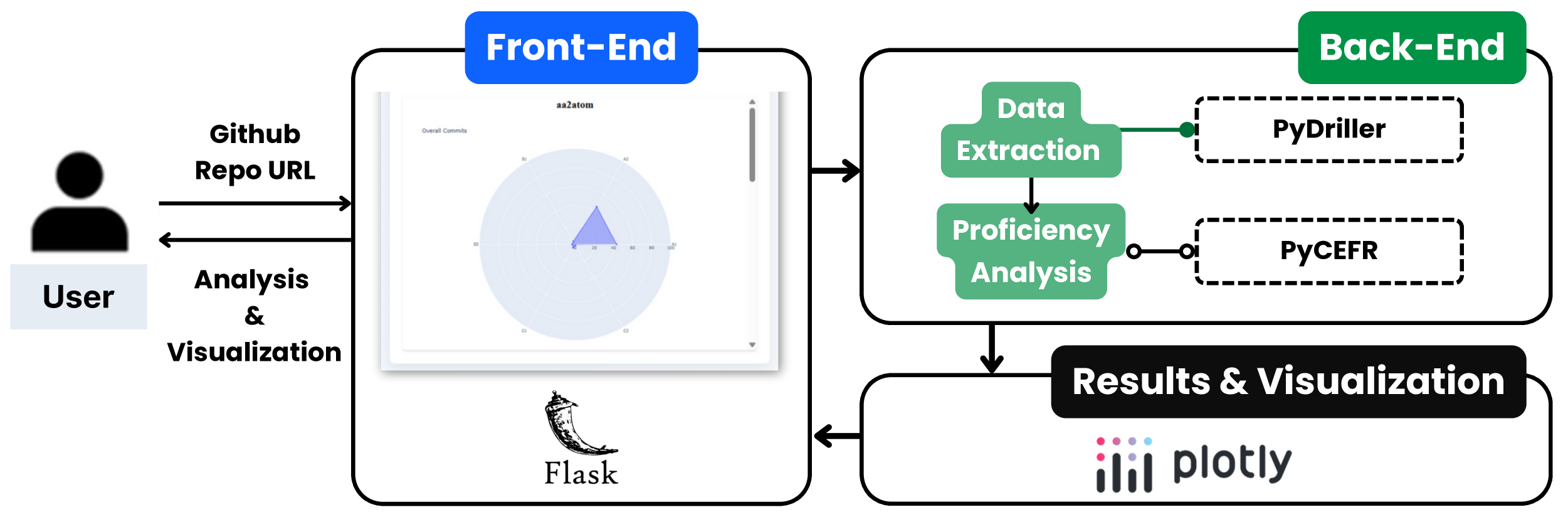}
    \caption{System architecture of PyGress}
    \label{fig:architecture}
\end{figure}

From \autoref{fig:architecture}, a user of PyGress gives a GitHub repository URL to the tool. Then, PyGress analysis starts by first cloning the given OSS project and passing it on to PyDriller~\cite{pydriller}, a Python library designed for exploring Git repositories and examining commit histories, to collect code change data based on each commit. PyDriller is configured to extract code changes in all Python files (\texttt{.py} extension) and commits data from those projects, recording them into the code changes database. Next, the pycefr tool is activated to extract the proficiency scores. Lastly, PyGress returns the visualizations of the proficiency data back to the users.

\subsection{Implementation}

PyGress follows a modular architecture that separates data processing, visualization, and user interaction. 
Our design of the PyGress system consists of three main modules:

\subsubsection{Back-end Module} The backend performs the core processing by employing PyDriller and pycefr to perform commit data extraction and Python proficiency analysis tasks.
    
\subsubsection{Front-end Module} A Flask-based web application enables users to submit GitHub repository links and explore analysis results through a web browser. This module dynamically loads the generated visualizations and shows them to the users.

\subsubsection{Visualization Module}
The front-end includes a visualization module, which processes the analysis results to generate interactive charts using Plotly. Two primary visual representations are provided. First, spider charts, which depict the distribution of proficiency levels for all contributors and each individual contributor. Second, slider graphs, which illustrate code proficiency changes throughout the project timeline. 

\begin{figure}[tb]
    \centering
    \includegraphics[width=0.7\columnwidth]{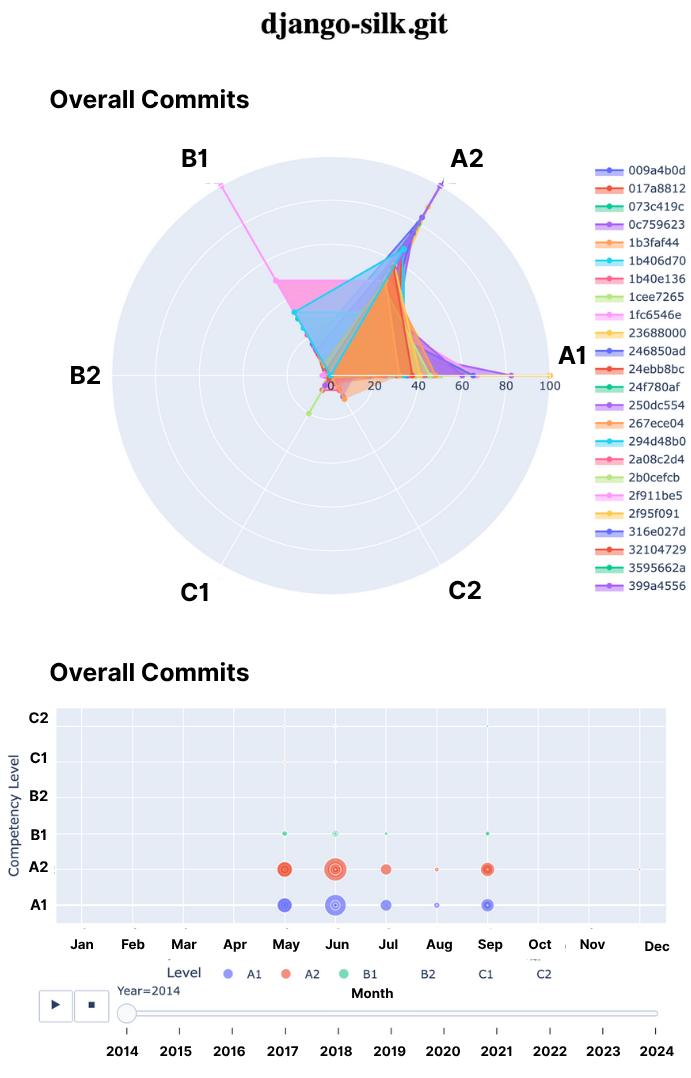}
    \caption{\textsf{django-silk}: Aggregated proficiencies--project level}
    \label{fig:screenshot-1}
\end{figure}

\subsection{Examples of PyGress Visualizations}
We chose the \textsf{django-silk} project as an example for the visualization of PyGress.
From Figure~\ref{fig:screenshot-1}, the first spider chart at the top shows the proficiency level of all the contributors in the project. We can observe that most of the Python code contributed to this project is at A1 and A2 proficiency levels, followed by B1, and some of B2, C1, and C2. Looking at the bottom chart of Figure~\ref{fig:screenshot-1}, the duration where most of the development activities occurred was during May--September 2014, when the project started. Most of the code committed during that period was at the A1, A2, and B1 levels. 

\begin{figure}[tb]
    \centering
    \includegraphics[width=0.7\columnwidth]{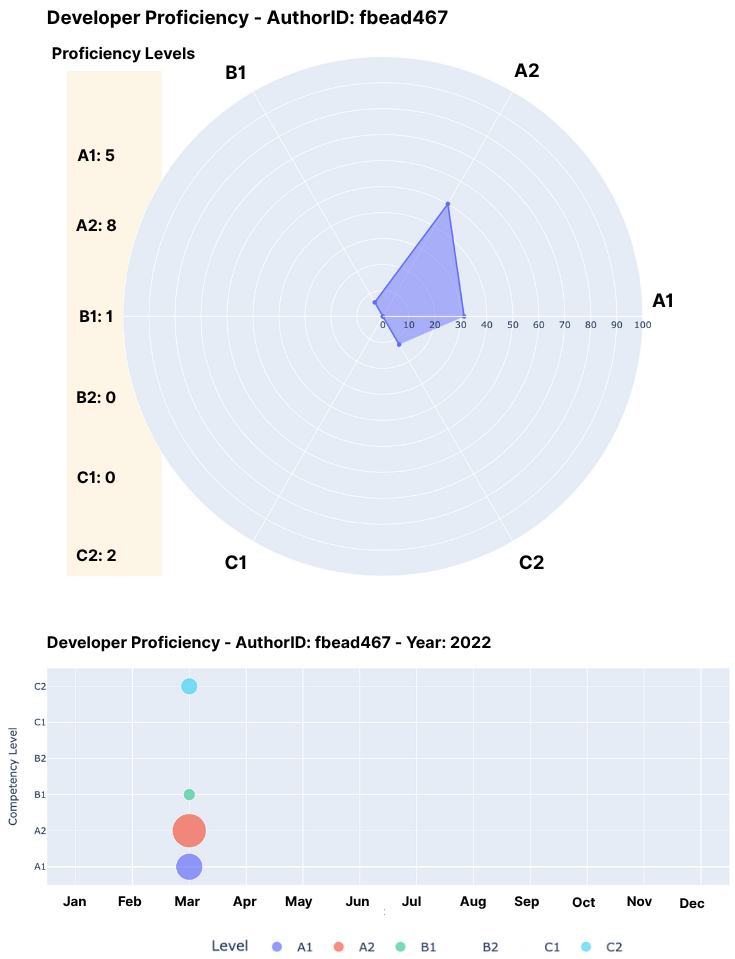}
        \caption{\textsf{django-silk}: Individual contributor's proficiency}
        \label{fig:screenshot-2}
\end{figure}

When investigating specifically at one contributor, \texttt{fbead467}\footnote{An automatically generated anonymized ID for preserving privacy. The tool can be configured to show the actual contributor's name or account name.} The result is as shown in Figure~\ref{fig:screenshot-2}. We can see from the spider chart that this contributor mostly contributed code at levels A1 and A2. However, he or she also committed some B1 and, importantly, C2, as well. This contributor committed the code in March 2022. Thus, the part of the C2 code that they committed may need to be carefully maintained since it may not be easy to understand for other contributors.

\begin{table*}[tb]
\caption{Python proficiency of 3 OSS projects}
\centering
\resizebox{0.97\textwidth}{!}{%
\begin{tabular}{lrrrrr rrrrrrr rrrrrrr}
\toprule
 & \multicolumn{6}{c}{django-silk} & \multicolumn{6}{c}{pandas-profiling} & \multicolumn{6}{c}{pytest-ansible} \\
 \cmidrule(lr){2-7} \cmidrule(lr){8-13} \cmidrule(lr){14-19}
Year & A1 & A2 & B1 & B2 & C1 & C2 & A1 & A2 & B1 & B2 & C1 & C2 & A1 & A2 & B1 & B2 & C1 & C2 \\
\midrule
2014 & 5,495 & 6,453 & 640 & 51 & \textbf{137} & \textbf{131} &  &  &  &  &  &  &  &  &  &  &  &  \\
2015 & 773 & 962 & 81 & 10 & \textbf{14} & \textbf{31} &  &  &  &  &  &  & 83 & 120 & 5 & 37 & 1 & 1 \\
2016 & 1,136 & 1,474 & 142 & 10 & \textbf{16} & \textbf{19} & 2,484 & 3,791 & 253 & 0 & 84 & 3 & 440 & 692 & 89 & 102 & 4 & 4 \\
2017 & 1,081 & 1,406 & 128 & 14 & \textbf{40} & \textbf{38} & 1,719 & 2,375 & 183 & 0 & 46 & 0 & 28 & 57 & 8 & 6 & 2 & 3 \\
2018 & 151 & 200 & 11 & 0 & 2 & 1 & 1,183 & 1,706 & 131 & 0 & 25 & 3 & 201 & 378 & 28 & 22 & 3 & 2 \\
2019 & 495 & 634 & 49 & 6 & 14 & 13 & 1,041 & 1,459 & 133 & 48 & 23 & 7 & 72 & 141 & 20 & 1 & 1 & 2 \\
2020 & 136 & 213 & 19 & 2 & 7 & 4 & 6,111 & 7,277 & 800 & 246 & \textbf{163} & \textbf{66} & 28 & 36 & 1 & 0 & 0 & 0 \\
2021 & 1,853 & 2,899 & 197 & 18 & 73 & 35 & 4,898 & 5,964 & 721 & 111 & \textbf{203} & \textbf{37} & 243 & 422 & 46 & 13 & 3 & 5 \\
2022 & 154 & 300 & 28 & 2 & 3 & 6 & 1,604 & 2,040 & 183 & 69 & \textbf{70} & \textbf{38} &  &  &  &  &  &  \\
2023 & 81 & 126 & 5 & 0 & 1 & 2 & 1,410 & 1,957 & 150 & 25 & \textbf{42} & \textbf{43} & 326 & 511 & 64 & 20 & 6 & 2 \\
2024 & 10 & 22 & 1 & 0 & 0 & 0 & 31 & 19 & 2 & 4 & 0 & 0 & 0 & 2 & 0 & 0 & 0 & 0 
\\
\midrule
Total & 11,365 & 14,689 & 1,301 & 113 & 307 & 280 & 20,481 & 26,588 & 2,556 & 503 & 656 & 197 & 1,421 & 2,359 & 261 & 201 & 20 & 19 \\
\bottomrule
\end{tabular}
}
\label{tab:oss_projects}
\end{table*}

\begin{table}[tb] 
    \centering
    \caption{Contributions of the most proficient contributor of each project}
    \label{tab:combined_proficiencies_c1_c2}
    \begin{tabular}{lllrr} 
        \toprule
        Project & Contributor & Year & C1 & C2 \\
        \midrule
        \multirow{2}{*}{django-silk} & \multirow{2}{*}{e721d399} & 2014 & 133 & 127 \\
        & & 2015 & 12 & 29 \\
        \midrule 
        \multirow{3}{*}{pandas-profiling} & \multirow{3}{*}{edf72917} & 2019 & 20 & 7 \\
        & & 2020 & 92 & 42 \\
        & & 2021 & 77 & 19 \\
        \midrule 
        \multirow{3}{*}{pytest-ansible} & \multirow{3}{*}{5ac7cd15} & 2016 & 4 & 4 \\
        & & 2017 & 2 & 3 \\
        & & 2018 & 3 & 2 \\
        \bottomrule
    \end{tabular}
\end{table}

\section{Practical Applications}
We applied PyGress to analyze the Python proficiency of three OSS projects, \textsf{django-silk}, \textsf{pandas-profiling}\footnote{https://github.com/pandas-profiling/pandas-profiling}, and \textsf{pytest-ansible}\footnote{https://github.com/ansible/pytest-ansible}. The result is shown in Table~\ref{tab:oss_projects}. We can see that the Python code in the three projects is mostly at the level A1 and A2. Looking only at the high proficiency code (i.e., C1 and C2) levels, we can observe that the \textsf{django-silk} contributors mostly committed high proficiency code more at the beginning of the project (2014--2017), potentially building the core logic of the project. In contrast, the contributors of \textsf{pandas-profiling} committed high proficiency code during the later years of the project (2020--2023). For \textsf{pytest-ansible}, there are not many high proficiency code constructs, and they are committed evenly across all the years. 

We further analyzed the contributors to find the one that contributed the most proficient code (C1 and C2) to each of the three projects, i.e., the \textit{most proficient contributor}. The result is shown in Table~\ref{tab:combined_proficiencies_c1_c2}. For \textsf{django-silk}, the most proficient contributor (e721d399) committed most proficient code during 2014--2015 (highlighted in bold text). For \textsf{pandas-profiling}, the most proficient contributor (edf72917) contributed the most proficient code during 2019--2021. Lastly, for \textsf{pytest-ansible}, the most proficient contributor (5ac7cd15) contributed the most proficient code during 2016--2018. These contributors can potentially be considered as a bus factor of the project due to their highly proficient code that may be difficult to understand by others.

\section{Conclusion}
We present an automated tool, PyGress, for evaluating Python code proficiency and its progression in OSS projects. The tool analyzes project commit history and visualizes the proficiencies for ease of understanding. The tool should be useful for the Python OSS project maintainers for getting insights into their codebase and managing the contributors accordingly for the sustainability of the project.
For future work, we plan to improve the processing speed of the PyGress tool by modifying pycefr to analyze the diff data. We also plan to add more visualizations, e.g., heatmaps, to show the code proficiency in each project's modules. Lastly, we plan to validate the tool with OSS project maintainers.

\section{Acknowledgement}
This research is partially supported by JSPS Kakenhi (A) JP24H00692 and the Faculty of ICT, Mahidol University.

\bibliographystyle{IEEEtran}
\bibliography{references}

\end{document}